\documentclass{article}
\pdfoutput=1
\usepackage{amsmath}
\usepackage{amssymb}
\usepackage{graphicx}
\usepackage{color}

\pdfpageheight\paperheight
\pdfpagewidth\paperwidth

\usepackage{jcappub}

\renewcommand{\[}{\begin{equation}}
\renewcommand{\]}{\end{equation}} 
\usepackage{array}

\begin{document}

\title{Signatures of  Horndeski gravity on the  Dark Matter Bispectrum}

\author[a]{Emilio Bellini,}

\author[b,a,c]{Raul Jimenez}

\author[b,a,d]{and Licia Verde}

\affiliation[a]{ICC, University of Barcelona, IEEC-UB, Mart\'i Franqu\'es, 1, E08028
Barcelona, Spain}

\affiliation[b]{ICREA (Instituci\'o Catalana de Recerca i Estudis Avan\c{c}ats)}

\affiliation[c]{Institute for Applied Computational Science, Harvard University,
MA 02138, USA}

\affiliation[d]{Institute of Theoretical Astrophysics, University of Oslo, 0315
Oslo, Norway}

\emailAdd{emilio.bellini@icc.ub.edu, raul.jimenez@icc.ub.edu, liciaverde@icc.ub.edu}

\abstract{We present a detailed study of second-order matter perturbations
for the general Horndeski class of models. Being the most general
scalar-tensor theory having second-order equations of motion, it includes
many known gravity and dark energy theories and General Relativity
with a cosmological constant as a specific case. This enables us to
estimate the leading order dark matter bispectrum generated at late-times
by gravitational instability. We parametrize the evolution of the first
and second-order equations of motion as proposed by Bellini and Sawicki (2014), where the free functions of the theory are assumed to be proportional to the dark energy density.
We show that it is unnatural
to have large $\gtrsim10\%$ ($\gtrsim  1\%$) deviations of the bispectrum
introducing even larger  $\sim30\%$ ($\sim 5\%$) deviations in the linear growth rate. Considering that measurements of the linear growth rate have much higher signal-to-noise than bispectrum measurements,
this indicates that for Horndeski models which reproduce the expansion
history and the linear growth rate as predicted by GR the dark matter
bispectrum kernel can be effectively modelled as the standard GR one.
On the other hand, an observation of a large bispectrum deviation
that can not be explained in terms of bias would imply either that
the evolution of perturbations is strongly different than the
evolution predicted by GR or that the theory of gravity is exotic (e.g., breaks the weak equivalence principle)
and/or fine-tuned.}

\maketitle

\section{Introduction}

One of the most challenging problems in modern cosmology is to understand
the nature of the accelerated expansion of the Universe at late-times
\cite{Perlmutter:1998np,Riess:1998cb}. The simplest model that explains
this behaviour is the $\Lambda$-cold dark matter one ($\Lambda\textrm{CDM}$),
where gravity is described by General Relativity (GR) at all scales
and the the cosmological constant $\Lambda$ is used in order to have
acceleration on cosmological scales. It is a simple model, with a
minimal number of parameters that fits exceptionally well a suite
of observations ranging from the early Universe to today and over
4 order of magnitude in scale \cite{Ade:2013zuv,Planck:2015xua}.
However, the value of $\Lambda$ appears to be too small to be explained
by fundamental physics (e.g.,~\cite{Weinberg:2000yb}).

One popular alternative is to add a scalar degree of freedom 
to the Einstein equations, either by acting as an additional and exotic
fluid of the system (as for Dark Energy (DE) models), or by modifying
directly the laws of gravity (as for Modified Gravity (MG) models).
In this paper we focus on the Horndeski theory introduced in \cite{Horndeski:1974wa,Deffayet:2011gz,Kobayashi:2011nu}.
The action reads

\begin{equation}
S=\int\mathrm{d}^{4}x\,\sqrt{-g}\left[\sum_{i=2}^{5}{\cal L}_{i}\,+\mathcal{L}_{\text{m}}[g_{\mu\nu}]\right]\,,\label{eq:action}
\end{equation}
where $g_{\mu\nu}$ is the metric, $\mathcal{L}_{\textrm{m}}$ is
the matter lagrangian, which contains just one single fluid composed
by Dark Matter (DM) and baryonic matter, and $\mathcal{L}_{i}$ are
\begin{eqnarray}
{\cal L}_{2} & = & K(\phi,\,X)\,,\\
{\cal L}_{3} & = & -G_{3}(\phi,\,X)\Box\phi\,,\\
{\cal L}_{4} & = & G_{4}(\phi,\,X)R+G_{4X}(\phi,\,X)\left[\left(\Box\phi\right)^{2}-\phi_{;\mu\nu}\phi^{;\mu\nu}\right]\,,\\
{\cal L}_{5} & = & G_{5}(\phi,\,X)G_{\mu\nu}\phi^{;\mu\nu}-\frac{1}{6}G_{5X}(\phi,\,X)\left[\left(\Box\phi\right)^{3}+2{\phi_{;\mu}}^{\nu}{\phi_{;\nu}}^{\alpha}{\phi_{;\alpha}}^{\mu}-3\phi_{;\mu\nu}\phi^{;\mu\nu}\Box\phi\right]\,.
\end{eqnarray}
Here, $K$, $G_{3}$, $G_{4}$ and $G_{5}$ are arbitrary functions
of the scalar field $\phi$ and its canonical kinetic term $X=-\phi^{;\mu}\phi_{;\mu}/2$.
The subscript $X$ represents a derivative with respect to (w.r.t.)
$X$, while a subscript $\phi$ would denote a derivative w.r.t.~the
scalar field $\phi$. The action, Eq.~(\ref{eq:action}), is the
most general action for a single scalar field that has second-order
equations of motion on any background and satisfies the weak equivalence
principle, i.e.,~all matter species are coupled minimally and
universally to the metric $g_{\mu\nu}$. It encompasses many of the
classical DE/MG models studied to explain the late-time cosmic acceleration:\footnote{Note that, since the Horndeski action, Eq.~(\ref{eq:action}), includes
both DE and MG models in a unified description, in the following we
will not distinguish between them. Indeed, we shall focus on the differences
between GR+$\Lambda$CDM and those models that include an extra scalar
degree of freedom (i.e.~DE/MG).} quintessence \cite{Ratra:1987rm,Wetterich:1987fm}, kinetic gravity
braiding \cite{Deffayet:2010qz,Kobayashi:2010cm,Pujolas:2011he}, galileons \cite{Nicolis:2008in,Deffayet:2009wt},
$f\left(R\right)$ \cite{Carroll:2003wy}. We must note that with
the Horndeski lagrangian it is not possible to describe models which
break the weak equivalence principle, such as coupled quintessence \cite{Amendola:1999er}
and non-universal disformal couplings \cite{Zumalacarregui:2012us,Koivisto:2012za,Zumalacarregui:2013pma},
or Lorentz-invariance-violating models (e.g.~Ho\v rava-Lifshitz gravity
\cite{Horava:2009uw,Blas:2009qj}), or beyond Horndeski theories \cite{Gleyzes:2014dya,Gao:2014soa}.

Measurements of the expansion history of the Universe are in excellent
agreement with $\Lambda\textrm{CDM}$, e.g.~\cite{Planck:2015xua}.
Then, it is justified to assume it as our fiducial model for the background
evolution. However, the same behaviour could be obtained assuming
a particular DE/MG model with the requirement that the equation of
state of the new degree of freedom is $w\left(t\right)\equiv\frac{p\left(t\right)}{\rho\left(t\right)}\thickapprox-1$,
where $\rho\left(t\right)$ and $p\left(t\right)$ are the background
energy density and pressure of the DE/MG fluid respectively.
 
 It is  important to look at the growth of cosmological perturbations
via gravitational instability as traced by the Large Scale Structure
(LSS) of the Universe to learn about the nature of the cosmic acceleration.
Indeed in $\Lambda\textrm{CDM}$, and in minimally coupled quintessence
models, once the expansion history is given, the growth of structures
is fully determined. This one-to-one correspondence is broken by the
additional degree of freedom if we use a generic sub-class of the Horndeski action,
Eq.~(\ref{eq:action}). As explained in greater detail in the following
section, it is therefore important to have a description that separates the
expansion history from the evolution of the perturbations. This result
can be achieved by using an Effective Field Theory approach, which
has been developed for Horndeski models in \cite{Jimenez:2011nn,Gubitosi:2012hu,Bloomfield:2012ff,Gleyzes:2013ooa,Gleyzes:2014qga,Gleyzes:2014rba,Bellini:2014fua}.

In this paper we look at the evolution of matter perturbations at
first and second orders by using the general Horndeski action Eq.~(\ref{eq:action}).
The first statistics of interest is the Power Spectrum (PS) $P\left(t,k\right)$
defined as
\begin{equation}
\left\langle \delta\left(t,\vec{k}_{1}\right)\delta\left(t,\vec{k}_{2}\right)\right\rangle \equiv\left(2\pi\right)^{3}\delta^{(3)}\left(\vec{k}_{1}+\vec{k}_{2}\right)P\left(t,k_{1}\right)\,,
\end{equation}
where $\delta\left(t,\vec{k}\right)\equiv\frac{\delta\rho\left(t,\vec{k}\right)}{\rho\left(t\right)}$
is the matter density contrast, $\delta^{(3)}$ is the 3-dimensional
Dirac delta, and $\left\langle \ldots\right\rangle $ indicates the
ensemble averaging over the possible configurations of the Universe.
The PS encodes all the statistical information of a Gaussian random
field. However, due to gravitational instability, which amplifies
the initial small fluctuations, the late Universe is highly non-Gaussian.
The lowest order statistics sensitive to the non-linearities is the
bispectrum $B\left(t,\vec{k}_{1},\vec{k}_{2},\vec{k}_{3}\right)$,
defined as
\begin{equation}
\left\langle \delta\left(t,\vec{k}_{1}\right)\delta\left(t,\vec{k}_{2}\right)\delta\left(t,\vec{k}_{3}\right)\right\rangle \equiv\left(2\pi\right)^{3}\delta^{(3)}\left(\vec{k}_{1}+\vec{k}_{2}+\vec{k}_{3}\right)B\left(t,\vec{k}_{1},\vec{k}_{2},\vec{k}_{3}\right)\,,\label{eq:bispectrumdefinition}
\end{equation}
where the Dirac delta imposes that only closed triangle configurations
are to be considered.

In GR the expansion history of the Universe determines the growth
of the perturbations at all orders. Therefore, linear and non linear
growth histories do not add qualitatively new information (beside
shrinking the statistical error-bars). In this paper we want to show
in which cases the DM bispectrum carries significantly new information
about the growth of structures that is not included in the linear-order
growth. With the data coming from current and forthcoming Large Scale
Structure (LSS) surveys, such as BOSS \cite{Eisenstein:2011sa}, WiggleZ
\cite{Drinkwater:2009sd}, DES \cite{Abbott:2005bi} or Euclid \cite{Laureijs:2011gra},
it will be possible to measure the PS and the bispectrum with unprecedented
statistical precision ($\sim1\%$). The PS is a well known tried and
tested statistical tool, a work-horse of any statistical analysis
of LSS, which is extremely high signal to noise, and with well understood
statistical properties. The linear growth rate can be measured in
several independent ways (e.g., weak lensing, redshift space distortions
etc.). On the other hand (mildly) non-linear growth is much harder
to measure. The most widespread tool is the bispectrum, but it has
much lower signal to noise and the information is spread out over
several configurations. It is complicated to model and time consuming
even in a standard GR-$\Lambda$CDM model (eg.~\cite{Gil-Marin:2014baa,Gil-Marin:2014sta}).
Then, we want to know to what extent the modelling developed for this "fiducial"
case (i.e.~GR-$\Lambda$CDM) is more general and can be applied more
widely.

The general result for the leading order DM bispectrum assuming Gaussian
initial conditions reads (e.g., see \cite{Bernardeau:2001qr} for
a review)
\begin{equation}
B\left(t,\vec{k}_{1},\vec{k}_{2},\vec{k}_{3}\right)=F_{2}\left(t,\vec{k}_{1},\vec{k}_{2}\right)\,P\left(t,k_{1}\right)\,P\left(t,k_{2}\right)+cyc.,\label{eq:bispectrumresult}
\end{equation}
where $cyc.$ contains the possible permutations w.r.t~$\vec{k}_{i}$
and $F_{2}$ is a model dependent kernel which was derived for Horndeski
models in \cite{Takushima:2013foa}. We shall show the expression
for $F_{2}$ in the next section. Here, it is important to note that
the functional form of the bispectrum in terms of the kernel and the
linear PS is identical to the standard one. As a result of Eq.~(\ref{eq:bispectrumresult}),
the tree-level DM bispectrum can be separated in terms of linear quantities,
i.e.~$P\left(t,\vec{k}\right)$, and second-order quantities, i.e.~$F_{2}\left(t,\vec{k}_{1},\vec{k}_{2}\right)$.
The DM bispectrum has been investigated using second-order perturbation
theory or simulations in simple DE/MG theories as Brans-Dicke/$f(R)$
or cubic galileon \cite{Bartolo:2013ws,Tatekawa:2008bw,GilMarin:2011xq,Bernardeau:2011sf,Borisov:2008xn}.
The result is that, in these Horndeski sub-models, the DM bispectrum
kernel appears to be close to the standard one within deviations at
the percent level, and then it can be approximated as $F_{2}\simeq F_{2\textrm{GR}}$.
The main goal of this paper is to investigate if it is possible to
generalise this approximation for a wider class of Horndeski-type
DE/MG models. In these cases, it would be possible to avoid the second-order
analysis, and concentrate \textit{just} on the evolution of linear
perturbations by applying directly the  bispectrum constraints obtained assuming $\Lambda $CDM/GR e.g., \cite{Gil-Marin:2014baa,Gil-Marin:2014sta}.

The paper is organised as follows. In Section \ref{sec:Approach} we
describe the approach we used and show the main equations, while Appendix
\ref{sec:Formulas} is reserved to useful formulas that are too long
to fit in the main text. In Section \ref{sec:Results} we present
and discuss the main results, and in Section \ref{sec:Conclusions}
we draw our conclusions.

\section{Approach\label{sec:Approach}}

We assume that the Universe  at large enough scales is well described by small
perturbations on top of a flat Friedmann-Robertson-Walker (FRW) metric.
We consider scalar perturbations in the Newtonian gauge, and  adopt the notation
of \cite{Ma:1995ey}. Then, up to second order the line element reads
\begin{equation}
\mathrm{d}s^{2}=-(1+2\Psi+\Psi^{(2)})\mathrm{d}t^{2}+a^{2}(t)(1-2\Phi-\Phi^{(2)})\mathrm{d}\boldsymbol{x}^{2}\,,
\end{equation}
where for brevity we have neglected the superscript $(1)$ for linear
perturbations. Here $\Psi$ and $\Phi$ are the metric perturbations and in this gauge they can be interpreted as the gauge invariant Bardeen's potentials \cite{Bardeen:1980kt}. The perturbation of the scalar field is rewritten
as $v_{X}\equiv-\delta\phi/\dot{\phi}$, where $\phi$ and $\delta\phi$
are the background and the perturbation of the scalar in Eq.~(\ref{eq:action})
respectively. In the following we will show the equations for the metric perturbations, after having decoupled them from the scalar field perturbations.

In \cite{Bellini:2014fua}, see also \cite{Bloomfield:2012ff,Gleyzes:2013ooa,Gubitosi:2012hu}
for an equivalent approach, the authors describe the evolution of
linear perturbations in general scalar-tensor theories belonging to
the Horndeski class of models, Eq.~(\ref{eq:action}). The idea is
to identify the minimum number of operators that fully specify the
linear evolution of those models (see \cite{Bellini:2014fua} for
details). These operators are independent of each other, i.e.,~they can
be parametrized independently. One of the advantages of this description
is that it separates the effects of the background expansion from
the effects of the perturbations. Then, the total amount of cosmological
information can be compressed into five functions of time plus one
constant, namely
\begin{equation}
\{\Omega_{\textrm{m}0},H\left(t\right),\alpha_{\textrm{K}}\left(t\right),\alpha_{\textrm{B}}\left(t\right),\alpha_{\textrm{M}}\left(t\right),\alpha_{\textrm{T}}\left(t\right)\}\,,\label{eq:firstorderlist}
\end{equation}
where $\Omega_{\textrm{m}0}$ is the value of the matter density fraction
today, $H\left(t\right)$ is the Hubble function and $\alpha_{i}\left(t\right)$
represent the linear freedom of the Horndeski class of models. $\alpha_{\textrm{K}}$,
the kineticity, is the most standard kinetic term present in simple
DE models as quintessence or k-essence. $\alpha_{\textrm{B}}$ is
called braiding, it represents a mixing between the kinetic terms
of the metric and the scalar. $\alpha_{\textrm{M}}$, the Planck mass
run rate, describes systems in which the Planck mass is varying with
time. $\alpha_{\textrm{T}}$ is the tensor speed excess, and it directly
modifies the speed of tensors. Both $\alpha_{\textrm{M}}$ and $\alpha_{\textrm{T}}$
produce anisotropic stress between the gravitational potentials $\Psi$
and $\Phi$, and they are signatures of modifications in the evolution
of the gravitational waves \cite{Saltas:2014dha}.  Generalizing
the results of \cite{Bellini:2014fua} to second-order in perturbation
theory we find four other functions of time which are free to vary
in the general Horndeski theory. However, for our purposes we just
need two of them, since the other two appear in front of terms that
are sub-dominant on sub-horizon scales ($k^{2}\gg a^{2}H^{2}$). Then,
the total freedom one can have is represented by
\begin{equation}
\{\Omega_{\textrm{m}0},H\left(t\right),\alpha_{\textrm{K}}\left(t\right),\alpha_{\textrm{B}}\left(t\right),\alpha_{\textrm{M}}\left(t\right),\alpha_{\textrm{T}}\left(t\right),\alpha_{5}\left(t\right),\alpha_{4}\left(t\right)\}\,,\label{eq:secondorderlist}
\end{equation}
where the new functions $\alpha_{5}$ and $\alpha_{4}$ represent
the second-order freedom of this theory and are defined in Appendix
\ref{sec:Formulas}. From their definitions, $\alpha_{5}$ and $\alpha_{4}$
can be considered as second-order corrections to $\alpha_{\textrm{M}}$
and $\alpha_{\textrm{T}}$, then we can expect that realistic DE/MG
theories will have 
\begin{equation}
\left|\alpha_{5},\alpha_{4}\right|\lesssim \left|\alpha_{\textrm{M}},\alpha_{\textrm{T}}\right|\lesssim {\cal O}(1)\,.\label{eq:naturalness}
\end{equation}
Models that do not satisfy the second inequality have a strong non-minimal
coupling, and they are expected to drastically modify the evolution
of first-order scalar and tensor perturbations. Then, they can be
excluded by observations that rely on linear theory, without the need
of second-order analysis.  The first inequality in Eq.~(\ref{eq:naturalness})
describes the fact that $\alpha_{5}$ and $\alpha_{4}$ are built
with the same Horndeski functions as $\alpha_{\textrm{M}}$ and $\alpha_{\textrm{T}}$,
i.e.,~$G_{4}$ and $G_{5}$ (see Eq.~(\ref{eq:alpha5}-\ref{eq:alpha6})
and Ref.~\cite{Bellini:2014fua} for comparison). In principle it
is possible to create a model in which the second-order functions
are larger than the first-order ones, but it would require fine tuning.
 We consider  Eq.~(\ref{eq:naturalness}) as a \textit{naturalness} condition,  which we will however not impose a priori.  While our analysis below is general and does not {\it assume} this hierarchy,  it is reasonable to concentrate attention on the "natural" regime of  Eq.  (\ref{eq:naturalness}).

As specified above, with this approach we separate the contribution of the
background from the perturbations. Indeed, the $\alpha_{i}$ functions
are constructed to be independent  form the background in the general
Horndeski class of models. Then, we can assume a flat $\Lambda\textrm{CDM}$
cosmology for the evolution of $H\left(t\right)$ and fix $\Omega_{\textrm{m}0}=0.31$
\cite{Planck:2015xua}, this being our fiducial model. The Friedmann and the conservation equations
read
\begin{align}
 & 2\dot{H}=-3H^{2}\tilde{\Omega}_{\text{m}}\\
 & \dot{\tilde{\Omega}}_{\text{m}}+3H\tilde{\Omega}_{\text{m}}\left(1-\tilde{\Omega}_{\text{m}}\right)=-H\tilde{\Omega}_{\text{m}}\alpha_{\textrm{M}}\,,\label{eq:backdensity}
\end{align}
where $\tilde{\Omega}_{\textrm{m}}\equiv\frac{\rho_{\textrm{m}}}{3M_{*}^{2}H^{2}}$
is the DM fractional density, and $M_{*}^{2}\left(t\right)$ the effective
Planck mass of the theory \cite{Bellini:2014fua}. Here, we do not
need to show explicitly the contribution of the energy density of
the DE fluid, since it has been eliminated using the Friedmann constraint,
i.e.~$\tilde{\Omega}_{\Lambda}=1-\tilde{\Omega}_{\text{m}}$. As
shown in \cite{Sawicki:2015zya}, if the sound speed of the DE/MG
fluid is sufficiently close to the speed of light ($\gtrsim0.1c$)
we can also assume the Quasi-Static approximation, in which time derivatives
are considered to be sub-dominant w.r.t.~space derivatives (i.e.~$a^{2}\ddot{\Phi}\ll k^{2}\Phi$).
In this regime it is possible to prove that $\alpha_{\textrm{K}}\left(t\right)$
does not enter the equations of motion. Finally, we choose a parametrization
for the remaining $\alpha_{i}$, precisely the one suggested in \cite{Bellini:2014fua}
\begin{equation}
\alpha_{i}\left(t\right)=\left(1-\tilde{\Omega}_{\textrm{m}}\right)c_{i}\,,\label{eq:parametrization}
\end{equation}
where $c_{i}$ are arbitrary constant. With this choice we ensure
the standard evolution of perturbations during radiation and matter
domination, while we allow for modifications when DE starts
dominating. This parametrization reproduces exactly the imperfect fluid behaviour on its tracking solution introduced in \cite{Deffayet:2010qz}, and it can describe general models in which the deviations w.r.t.~$\Lambda$CDM are smooth. Then, Eq.~(\ref{eq:secondorderlist}) reduces to\footnote{Note that in the literature the simbol $c_{\textrm{T}}$ often identifies the speed of tensors. In this paper we use it only as the parameter that describes $\alpha_{\textrm{T}}$ in our parametrization, Eq.~(\ref{eq:parametrization}).}
\begin{equation}
\{c_{\textrm{B}},c_{\textrm{M}},c_{\textrm{T}},c_{5},c_{4}\}\,.\label{eq:finallist}
\end{equation}
Note that the naturalness condition Eq.~(\ref{eq:naturalness}) then becomes:
 \begin{equation}
\left | c_{5},c_{4}\right|\lesssim\left| c_{\textrm{M}},c_{\textrm{T}}\right|\lesssim {\cal O}(1)\,.\label{eq:naturalnessC}
\end{equation}
Before evaluating the linear and second-order evolution, we will adopt
a further simplification. It has been noticed in \cite{Bellini:2014fua}
the existence of the braiding scale
\begin{equation}
\frac{k_{\text{B}}^{2}}{a^{2}H^{2}}=\frac{9}{2}\tilde{\Omega}_{\textrm{m}}+2\left(\frac{3}{2}+\frac{\alpha_{\textrm{K}}}{\alpha_{\textrm{B}}^{2}}\right)\left(\alpha_{\textrm{M}}-\alpha_{\textrm{T}}\right)\,.\label{eq:braidingscale}
\end{equation}
This represents the only new scale in the linear equations beyond
the usual Jeans length. Models that evolve outside this scale have a perfect fluid structure, while clustering DE on
linear scales is possible well inside the braiding scale. In the following we focus only on two classes of models,
the ones that have sub-braiding evolution ($\alpha_{\textrm{B}}^{2}\gg\alpha_{\textrm{K}}$),
and the ones with super-braiding evolution ($\alpha_{\textrm{B}}^{2}\ll\alpha_{\textrm{K}}$).
As we shall see, in this limit we ensure that the linear-order DM
density evolution is scale independent. In the intermediate regime,
where $\alpha_{\textrm{B}}^{2}\sim\alpha_{\textrm{K}}$, we expect
the growth of perturbations to be scale dependent. This would modify
the shape of the PS, and these models in this regime could be already
studied at this level without including the bispectrum. In fact the
shape of the (matter) PS at linear and mildly non-linear scales is a high signal to noise and relatively robust quantity to measure. It is much less affected by e.g.,~ non-linear galaxy bias than measurements of  e.g., the growth rate of structures.

\subsection{Linear-order and second-order evolution}

The linearized Einstein equations together with the equation of motion
for the scalar field, as usual, provide the Poisson equations for
the gravitational potentials, e.g.~\cite{DeFelice:2011hq,Amendola:2012ky,Motta:2013cwa},
\begin{align}
 & \frac{k^{2}}{a^{2}}\Psi=-\frac{3}{2}H^{2}\tilde{\Omega}_{\textrm{m}}G_{\Psi}\delta_{\textrm{m}}\label{eq:Poissonpsi}\\
 & \frac{k^{2}}{a^{2}}\Phi=-\frac{3}{2}H^{2}\tilde{\Omega}_{\textrm{m}}G_{\Phi}\delta_{\textrm{m}}\,,\label{eq:Poissonphi}
\end{align}
where $\delta_{\textrm{m}}\equiv\delta\rho_{\textrm{m}}/\rho_{\textrm{m}}$
is the density contrast, while $G_{\Psi}\left(t\right)$ and $G_{\Phi}\left(t\right)$
are the effective Newton's constants. In GR we have $G_{\Psi}=G_{\Phi}=1$,
while their definition in our case is given in Appendix \ref{sec:Formulas}. It is interesting to note that $G_{\Psi}$ and $G_{\Phi}$ are scale independent in the limits we are considering, i.e.~sub-braiding and super-braiding. Eqs.~(\ref{eq:Gpsi}-\ref{eq:Gphi}) are presented in Appendix \ref{sec:Formulas} in the sub-braiding case. Their expression in the opposite limit is identical to Eqs.~(\ref{eq:Gpsi}-\ref{eq:Gphi}) imposing $\alpha_{\textrm{B}}\rightarrow 0$.
Note that, in order to obtain the previous equations, we already decoupled
the metric from the scalar field perturbations $v_{X}$. Its effects
are fully encoded in $G_{\Psi}$ and $G_{\Phi}$. In addition, the
conservation of the stress-energy tensor gives the linear evolution
of the DM density contrast, which reads
\begin{equation}
\ddot{\delta}_{\textrm{m}}+2H\dot{\delta}_{\textrm{m}}=\frac{1}{2}G_{\Psi}\tilde{\rho}_{\textrm{m}}\delta_{\textrm{m}}\,.\label{eq:diffdelta1}
\end{equation}

Following the same procedure described in \cite{Bartolo:2013ws},
we can get the evolution of the second-order density fluctuations.
The result is an equation which is identical to Eq.~(\ref{eq:diffdelta1}),
with the addition of a non-vanishing source ($\tilde{S}_{\delta}$)
composed by products of first-order quantities (e.g.~$\delta_{\textrm{m}}\delta_{\textrm{m}}$),

\begin{equation}
\ddot{\delta}_{\textrm{m}}^{(2)}+2H\dot{\delta}_{\textrm{m}}^{(2)}-\frac{1}{2}\tilde{\rho}_{\textrm{m}}G_{\Psi}\delta_{\textrm{m}}^{(2)}=\tilde{S}_{\delta}\,.\label{eq:diffdelta2}
\end{equation}
The exact form of $\tilde{S}_{\delta}$ is not important for the purpose
of this paper. More details can be found in \cite{Takushima:2013foa},
where the bispectrum for the Horndeski class of models was first studied.
The solution of Eq.~(\ref{eq:diffdelta2}) is given by
\begin{align}
\delta_{\textrm{m}}^{(2)}\left(t,\vec{k}\right) & =\int\frac{d^{3}q_{1}d^{3}q_{2}}{(2\pi)^{3}}\delta_{D}\left(\vec{k}-\vec{q}_{1}-\vec{q}_{2}\right)F_{2}\left(t,\vec{q}_{1},\vec{q}_{2}\right)\delta_{\textrm{m}}(t,\vec{q}_{1})\delta_{\textrm{m}}(t,\vec{q}_{2})\,,\label{eq:soldelta2}
\end{align}
where the kernel $F_{2}$ coincides with the kernel shown in Eq.~(\ref{eq:bispectrumresult})
and encodes the second-order modifications w.r.t.~the usual Newtonian
one \cite{Bernardeau:2001qr}. Here we have  identified the evolution of the second-order DM perturbations with a particular solution of Eq.~(\ref{eq:diffdelta2}), neglecting the homogeneous part. The homogeneous solution contains both a possible primordial non-Gaussianity and a non-primordial contribution. Since in our analysis the evolution of cosmological perturbations is standard up to redshift $z\sim1$ by construction, the non-primordial contribution remains subdominant on mildly non-linear scales, see e.g.~\cite{Bartolo:2005xa,Bartolo:2006fj}. 
In order to select the leading-order
contribution on sub-horizon scales, we assume Gaussian initial conditions and we perform an expansion in terms
of $\epsilon\equiv\frac{aH}{k_{i}}$ in $F_{2}$, where $k_{i}$ stands
for $k$, $q_{1}$ and $q_{2}$. One could argue that with this method
we are neglecting terms proportional to $\frac{aH}{q_{i}}$, which
are important in the integral when $q_{i}\rightarrow0$. However,
these terms represent wavelengths that are super-horizon, they can
be considered as a backreaction and reabsorbed into the background
evolution. Then, the kernel $F_{2}$ read \cite{Takushima:2013foa}
\begin{equation}
F_{2}\left(t,\vec{q}_{1},\vec{q}_{2}\right)=C\left(t\right)+\frac{\left(q_{1}^{2}+q_{2}^{2}\right)\mu}{q_{1}q_{2}}-\left(1-\frac{1}{2}C\left(t\right)\right)\left(1-3\mu^{2}\right)\,.\label{eq:kernel}
\end{equation}
The full definition for the function $C\left(t\right)$ is given in
Appendix \ref{sec:Formulas} in Eqs.~(\ref{eq:C}-\ref{eq:c}), and
it represents the only effect of Horndeski type of DE/MG at the leading
order on mildly non-linear scales. After some algebra, it is possible to demonstrate that our results, Eq.~(\ref{eq:kernel}) and Eqs.~(\ref{eq:C}-\ref{eq:c}), reproduce the results of \cite{Takushima:2013foa}. With our notation, the standard
value for this function in an Einstein de Sitter Universe is $C\left(t\right)=34/21$.
The dependence of this value on the cosmology is very weak. Indeed
in GR, for $\Lambda\textrm{CDM}$ and minimally coupled models as quintessence, the corrections to
the standard value are negligible \cite{Scoccimarro:2000ee,Catelan:1994kt,Scoccimarro:2000sn,Bouchet:1992uh,Bernardeau:1993qu}.
Note that the denominator of Eq.~(\ref{eq:c}) does not produce divergences.
Indeed, after some algebra it is possible to demonstrate that, if
we tune $\alpha_{\textrm{B}}+2\alpha_{\textrm{M}}-\alpha_{\textrm{T}}\left(2-\alpha_{\textrm{B}}\right)=\mathcal{O}\left(\epsilon\right)$,
the numerator will suppress these divergences with factors $\mathcal{O}\left(\epsilon^{2}\right)$.
Even if Eqs.~(\ref{eq:C}-\ref{eq:c}) appear long and complicated,
it is possible to identify three main objects. The first one depends
only on first-order quantities, the second depends on $(\alpha_{5},\dot{\alpha}_{5})$,
and the third on $(\alpha_{4},\dot{\alpha}_{4})$. Then, using our
parametrization  for the Horndeski functions $\alpha_i$, Eq.~(\ref{eq:parametrization}), we can rewrite $C(t)$ in the expression for the second order gravitational kernel  Eq.~(\ref{eq:kernel}) as
\begin{equation}
C\left(t\right)=A_{0}\left(t\right)+A_{5}\left(t\right)c_{5}+A_{4}\left(t\right)c_{4}\,,\label{eq:A0A5A6}
\end{equation}
 which can be explicitly seen by using Eq.~(\ref{eq:parametrization}) in the expressions of Eqs.~(\ref{eq:C}-\ref{eq:c}). 
Here, all the functions $\{A_{0},A_{5},A_{4}\}$ depend only on background
and first-order quantities. 
Note that  $\alpha_5$ and $\alpha_{4}$, and therefore in our parameterisation   $c_5$ and $c_{4}$, are zero in GR, and in popular MG models such as Brans-Dicke/$f(R)$, kinetic gravity braiding, cubic galileons.

This means that, once the expansion history
and the linear evolution are fixed, the only freedom we have is in
the choice of two parameters $\{c_{5},c_{4}\}$. In the next section
we show the magnitude of $\{A_{0},A_{5},A_{4}\}$ as a function of
$\{c_{\textrm{B}},c_{\textrm{M}},c_{\textrm{T}}\}$ in order to clarify
 under which conditions it is possible to have large deviations  from the GR kernel in the DM bispectrum.
Indeed, this analysis  will show the minimum value for $\{c_{5},c_{4}\}$
needed in order to substantially modify Eq.~(\ref{eq:kernel}) and
finally the DM bispectrum, Eq.~(\ref{eq:bispectrumresult}).

\section{Results\label{sec:Results}}

In this section we show the main results of this paper. At linear-order
we focus on the effective Newton's constant $G_{\Psi}$, Eq.~(\ref{eq:Gpsi}),
the growth rate $f\equiv\frac{d\ln\delta_{\textrm{m}}}{d\ln a}$ and
the slip parameter $\bar{\eta}\equiv\frac{2\Psi}{\Psi+\Phi}$. The
Newton's constant measures the strength of the gravitational force,
and modifies directly the matter PS. The growth rate tracks the growth
of linear perturbations. The slip parameter quantifies the anisotropic
stress between the two gravitational potentials $\Psi$ and $\Phi$,
and can be a signature of non-minimal coupling between the metric
and the scalar field \cite{Saltas:2014dha}. It can be measured by comparing weak lensing
with redshift space distortions and the estimated errors with data
from future surveys such as Euclid are $\sim10\%$ in case it is not
scale dependent \cite{Amendola:2013qna}. All these quantities are
indications of deviations w.r.t.~the prediction of a pure $\Lambda\textrm{CDM}$,
which we consider as our reference model. From Eqs.~(\ref{eq:Poissonpsi}-\ref{eq:Poissonphi}-\ref{eq:diffdelta1})
it is possible to note that they are independent each other. With
present surveys   $f$ is measured with $\sim 6\%$  precision \cite{Beutler:2013yhm,Samushia:2013yga} and  $G_{\Psi}$ is measured (although in a model-dependent way) with $\sim 10\%$ precision \cite{Amendola:2003eq} but $\bar{\eta}$ is currently poorly constrained ($\sim 100\%$ error) e.g.,\cite{Daniel:2009kr}.

Future data
will increase this precision to  roughly one order of magnitude. Then, 
 for viable models we impose that the  most stringent condition
\begin{equation}
\left|\frac{f}{f_{\Lambda\textrm{CDM}}}-1\right|\lesssim6\%\,.\label{eq:firstordercondition}
\end{equation}
is satisfied.   Should future surveys not find any deviations from standard cosmology, this condition will become much stronger, with deviations allowed only  below the 1\%.   We explore the parameter space $\{c_{\textrm{B}},c_{\textrm{M}},c_{\textrm{T}}\}$
and we ensure that the models we pick do not suffer from gradient
and ghost instabilities, both on the scalar and the tensor sectors
(see \cite{Bellini:2014fua} for details).
For understanding the behaviour of the   second-order  bispectrum we study
the quantities $\{A_{0},A_{5},A_{4}\}$ evaluated today and defined
in Eq.~(\ref{eq:A0A5A6}) as a function of the free parameters. Indeed
they are the only modifications to  the kernel of second-order
DM matter perturbations. We choose to estimate them today in order
to maximise the effects of DE/MG on the bispectrum. With current surveys
it is possible to estimate the bispectrum with a $\sim10\%$ precision,
due to statistics and systematic uncertainties  \cite{Gil-Marin:2014baa,Gil-Marin:2014sta}. With
future surveys it will be possible to reduce the statistical uncertainties,
and, optimistically reach a $\sim1\%$ precision  and accuracy when averaged over all bispectrum configurations.
Then, in the following we consider bispectrum kernels modified by
$\sim1\%$ or less as kernels that are close to the standard result. On the
contrary, models that can show interesting signatures of DE/MG on
mildly non-linear scales with the next generation of experiments are
those models with modifications of the bispectrum kernel $\gtrsim10\%$.

\subsection{Small braiding}

In this section we consider models with small braiding, i.e.~$\alpha_{\textrm{B}}\rightarrow0$.
This ensures that the scalar degree of freedom evolves always on super-braiding
scales. In this case the scalar field has a perfect fluid structure,
with the possibility of having a non-minimal coupling given by $\{c_{\textrm{M}},c_{\textrm{T}}\}$.
However, after some algebra one can show that the dependence on $c_{\textrm{T}}$
vanishes both at first and at second orders. This result is a consequence
of choosing $\Lambda\textrm{CDM}$ for the background and $\alpha_{\textrm{B}}=0$,
it is therefore independent of the parametrization for the $\alpha_i$ functions we chose. Then,
at first-order we are left with just one parameter to vary, i.e.~$c_{\textrm{M}}$.

\begin{figure}
\centering\includegraphics[scale=0.8]{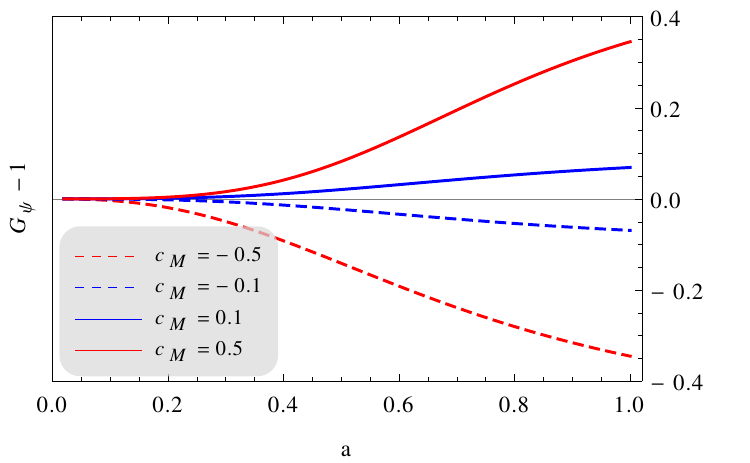}\\
\includegraphics[scale=0.8]{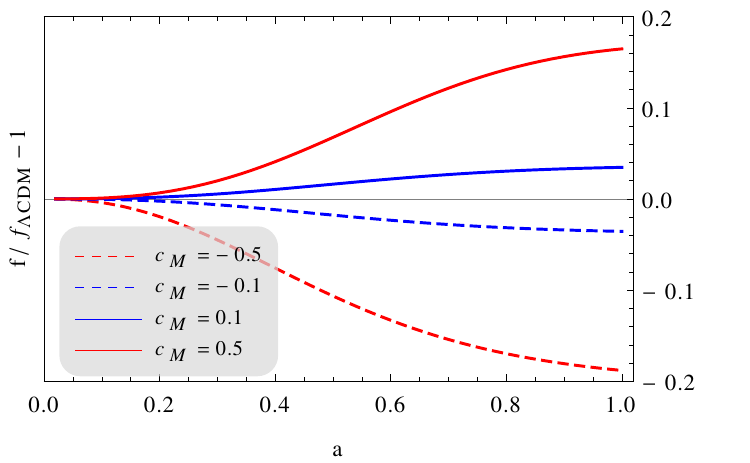}\\
\includegraphics[scale=0.8]{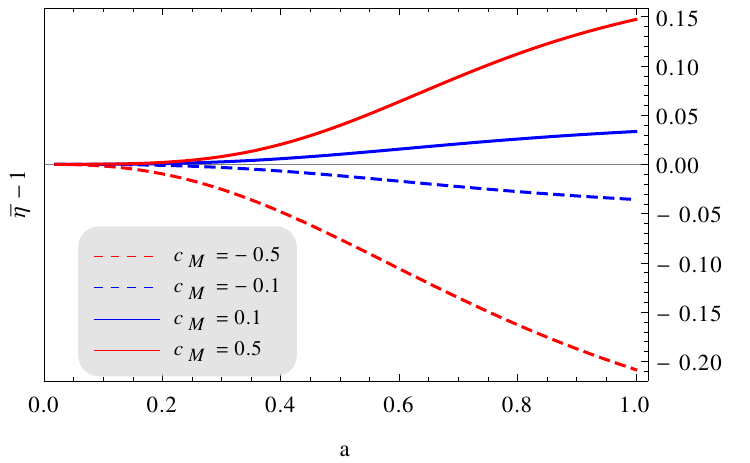}\protect\caption{Evolution of the effective Newton's constant $G_{\Psi}$ (top panel),
the linear growth rate $f$ (central panel) and the slip parameter
$\bar{\eta}$ (bottom panel) as a function of the scale factor $a$.
In all the panels we take the relative deviations w.r.t.~our fiducial
model, i.e.~$\Lambda\textrm{CDM}$. Starting from the bottom lines,
the different models are $c_{\textrm{M}}=-0.5$ (red dashed line),
$c_{\textrm{M}}=-0.1$ (blue dashed line), $c_{\textrm{M}}=0.1$ (blue
solid line) and $c_{\textrm{M}}=0.5$ (red solid line).\label{fig:noBraidingfirst}}
\end{figure}

In Fig.~(\ref{fig:noBraidingfirst}) we show the evolution of the
effective Newton's constant $G_{\Psi}$ top panel), the growth rate
$f$ (central panel) and the slip parameter $\bar{\eta}$ (bottom panel)
as a function of the scale factor. Increasing the value of $\left|c_{\textrm{M}}\right|$
is equivalent to increase the deviations w.r.t.~the standard values
$G_{\Psi}=1$, $f=f_{\Lambda\textrm{CDM}}$ and $\bar{\eta}=1$. Note that the extreme models, $c_M=\pm 0.5$, are already  disfavoured from current data as they  induce effects on observable quantities that are larger than the limit of  Eq.~(\ref{eq:firstordercondition}). Thus $c_M \lesssim 0.2$ but for forecasted errors from future surveys, should there be no deviations from standard cosmology, then   $c_M \ll 0.1$.

\begin{figure}
\centering\includegraphics[scale=0.6]{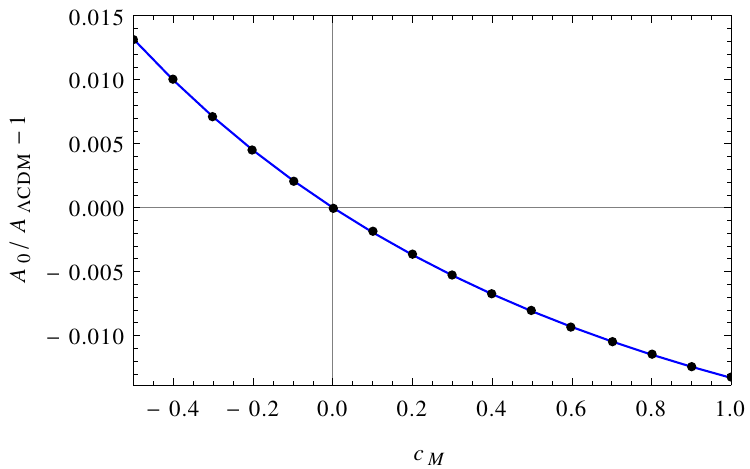}\includegraphics[scale=0.6]{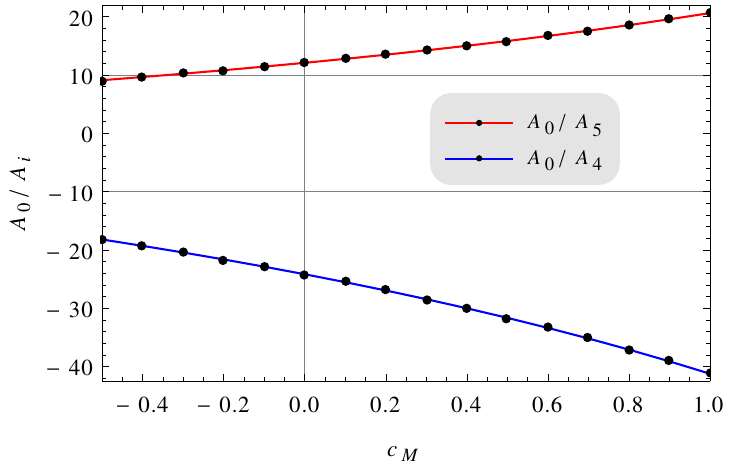}\protect\caption{In the left panel we show the relative deviation of $A_{0}$ w.r.t.~$A_{\Lambda\textrm{CDM}}$,
i.e.~$A_{0}$ for a pure $\Lambda\textrm{CDM}$, for different models.
In the right panel $A_{0}/A_{5}$ (red line) and $A_{0}/A_{4}$ (blue
line). In both panels the quantities are calculated today as a function
of $c_{\textrm{M}}$. We vary the parameter $c_{\textrm{M}}$ from
$-0.5$ to $1$, since the upper and the lower values already show
significant deviations at first-order (e.g.~Fig.~\ref{fig:noBraidingfirst}).
The left panel shows that the standard term of the kernel, Eq.~(\ref{eq:kernel}),
have deviations w.r.t.~the GR one $\lesssim1\%$ for all the cases
under consideration. The right panel shows that $c_{5}$ and $c_{4}$
should be $\gtrsim1$ in order to have an interesting effect on the
bispectrum kernel. \label{fig:noBraidingsecond}}
\end{figure}

In Fig.~(\ref{fig:noBraidingsecond}) we show the second-order interesting
quantities. In particular, in the left panel we plot the value of
$A_{0}(a=1)$, Eq.~(\ref{eq:A0A5A6}), normalized by its value in
a $\Lambda\textrm{CDM}$ universe as a function of the parameter $c_{\textrm{M}}$.
One can see that for small values of $c_{\textrm{M}}$ the standard
limit is recovered, while one can reach deviations of the order of
$1\%$ when $c_{\textrm{M}}=1$. However such large value for $c_M$ induces uncomfortably large changes on first-order quantities, see discussion around   Fig.~(\ref{fig:noBraidingfirst}). We can conclude
that $A_{0}$ in Eq.~(\ref{eq:A0A5A6}) is  effectively standard.
Therefore the only models that  have a chance to significantly change the bispectrum are those where $\alpha_5,\alpha_{4}$  are non-zero. 

 In the right panel we plot the values of $A_{0}/A_{5}$ and $A_{0}/A_{4}$,
Eq.~(\ref{eq:A0A5A6}), calculated today as a function of $c_{\textrm{M}}$.  In the relevant range for $c_{\textrm{M}}$, $A_5$ is 10 to 20 times smaller than $A_0$ and $|A_{4}|$ is 20 to 40 times smaller than $A_0$. 
The ratio $A_{0}/A_{i}$ therefore indicates how large the parameters $\{c_{5},c_{4}\}$
have to be in order to modify the bispectrum kernel, Eq.~(\ref{eq:kernel}).
The result is that, in the limit $c_{\textrm{M}}\rightarrow0$ or
when it is negative we need $\{c_{5},c_{4}\}\gtrsim\{1,2\}$  in order
to have a kernel modified by $10\%$  and  $\{c_{5},c_{4}\} > \{0.1,0.2\}$ to have a kernel modified by more than 1\%. When $c_{\textrm{M}}\rightarrow1$,
we need even larger values of $\{c_{5},c_{4}\}$ to have the same
deviation.

In conclusion, for models with negligible braiding, it is necessary
to require $c_{\textrm{M}}\ll1$ in order to have the evolution of
the linear perturbations not in tension with current constraints,
Eq.~(\ref{eq:firstordercondition}) \cite{Beutler:2013yhm,Samushia:2013yga}.
As shown in Fig.~\ref{fig:noBraidingsecond}, at second-order the
kernel, Eq.~(\ref{eq:kernel}), is weakly dependent on the value
of $c_{\textrm{M}}$ and its modifications are mostly due to the parameters
$c_{5},c_{4}$. It is possible to note that the kernel is modified
by less than $1\%$ w.r.t.~the standard one if we impose $c_{5},c_{4}\lesssim 0.1$.
Differences of order $10\%$  can arise  for models with $c_{5},c_{4}\gtrsim1$. Then, an interesting bispectrum
kernel modification can happen only if the parameters of the DE/MG
model we consider do not satisfy the expected natural hierarchy, Eq.~(\ref{eq:naturalnessC}).

\subsection{Large braiding}

Here we want to study models that evolve on sub-braiding scales. For
these models we expect a more interesting physics, since the braiding
causes the DE 
to cluster on linear scales. On the other hand,
the equations are longer and there are no reduction on the linear-order
free parameters. We then explore the parameter space $\{c_{\textrm{B}},c_{\textrm{M}},c_{\textrm{T}}\}$
using two grids of 30 points in each direction. As we shall see, the
dependence of the interesting first and second-order functions on
the parameters is smooth, then the choice of using just 30 points
is justified. For generic DE/MG models falling in our description,
the expected values for the free parameters are naturally $\mathcal{O}\left(1\right)$.
Then, for the first grid we assumed $c_{\textrm{B}}\in\left[-3;3\right]$, $c_{\textrm{M}}\in\left[-1.5;1.5\right]$
and $c_{\textrm{T}}\in\left[-0.75;0.75\right]$.  We used the second grid to focus on those models that have parameters closer to their standard value  (i.e.~zero): $c_{\textrm{B}}\in\left[-1;1\right]$, $c_{\textrm{M}}\in\left[-0.5;0.5\right]$
and $c_{\textrm{T}}\in\left[-0.5;0.5\right]$. With $c_{\textrm{M}}$,
and in particular with $c_{\textrm{T}}$, we have been more restrictive
than with $c_{\textrm{B}}$, since they regulate the non-minimal coupling
of the theory and they have a bigger effect on the evolution of perturbations.
Thus, our bounds can be considered representative of the full parameter
space. Here, it is important to recall that the kineticity, $\alpha_{\textrm{K}}$, does not appear in the QS equations, Eqs.~(\ref{eq:Poissonpsi}-\ref{eq:Poissonphi}). Then, it is always possible to tune the coefficient $c_{\textrm{K}}$ in order to satisfy the large braiding condition, i.e.~$\alpha_{\textrm{B}}^{2}\gg\alpha_{\textrm{K}}$, without modifying our results. For each point we check the stability conditions, we calculate
the value for the fractional deviation of the growth rate today w.r.t.~$\Lambda\textrm{CDM}$,
i.e.~$f\left(a=1\right)/f_{\Lambda}\left(a=1\right)-1$, the effective
Newton's constant, i.e.~$G_{\Psi}\left(a=1\right)-1$, and the slip
parameter, i.e.~$\bar{\eta}\left(a=1\right)-1$. We then estimate
the values of the functions $A_{0}/A_{\Lambda}$, $A_{0}/A_{5}$ and
$A_{0}/A_{4}$.

If we require that the  linear order quantity $f$ should not be  modified
by more than $6\%$ w.r.t.~standard gravity, Eq.~(\ref{eq:firstordercondition}), we find that  the second-order parameters should be at least
$c_{5},c_{4}\gtrsim2 (1)$ in order to have an effect of at least $10\% (3\%)$
on the kernel Eq.~(\ref{eq:kernel}).

If we want to keep $c_{5},c_{4}$
small ($\lesssim1$) and have large modifications on the kernel $\sim10\%$ ($3\%$) 
we have to allow for deviations of $\sim30\%$ ($10\%$) at linear-order. However, to  find signatures of 
these kind of models, a higher-order correlations  (and thus second-order) analysis is not needed, since
any signature of DE/MG could be seen easily using observables that
rely on linear theory. There are few exceptions to this result in the region of the parameter space where $c_{\textrm{T}}\sim -0.5$. Here we can keep $c_{5},c_{4}$ small ($\lesssim1$), have deviations on the linear growth rate $\lesssim 6\%$ and have large deviations on the bispectrum kernel ($\sim 10\%$). However, such $c_{\textrm{T}}$ values imply  a speed of tensors significantly smaller than the speed of light  which can be ruled out by
observations of ultra-high energy cosmic rays \cite{Moore:2001bv,Elliott:2005va,Kimura:2011qn}.

\begin{figure}
\centering\includegraphics[scale=0.42]{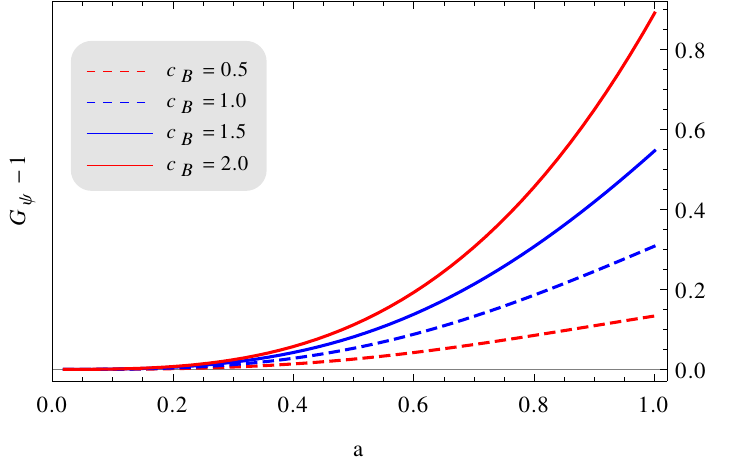}\includegraphics[scale=0.42]{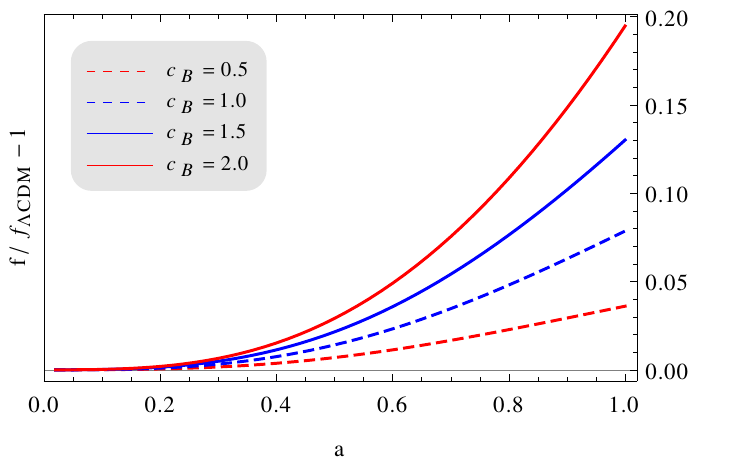}\includegraphics[scale=0.42]{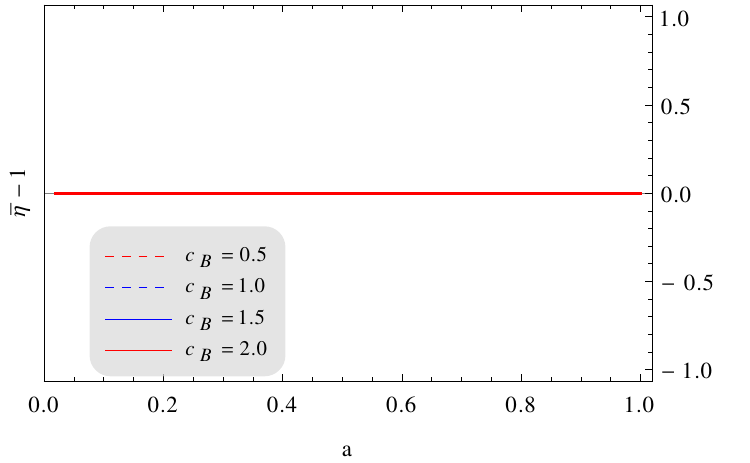}\\
\includegraphics[scale=0.42]{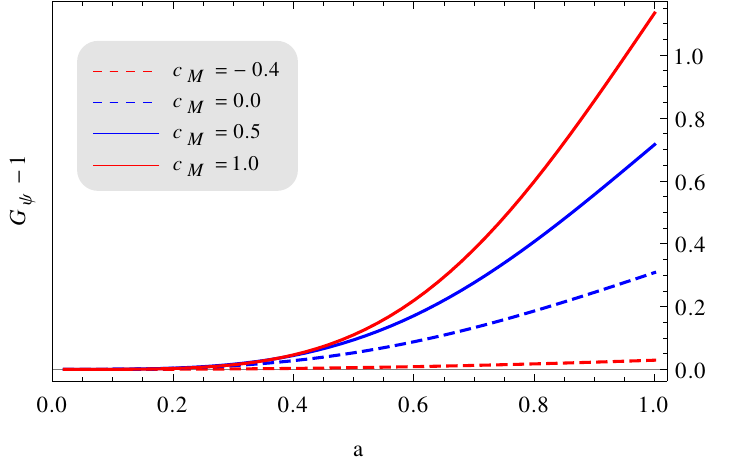}\includegraphics[scale=0.42]{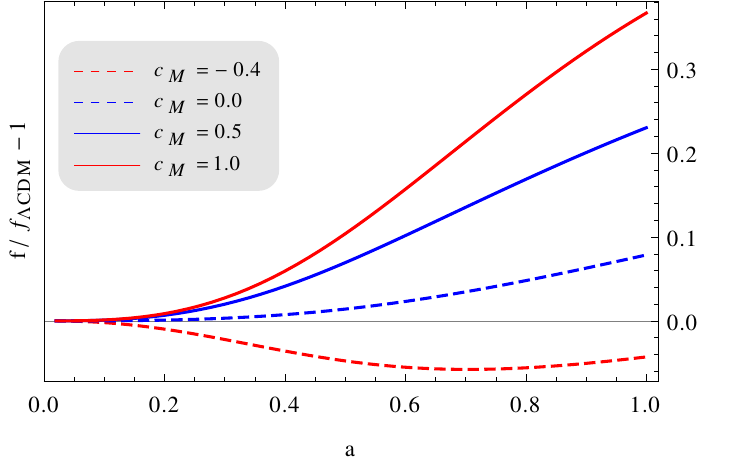}\includegraphics[scale=0.42]{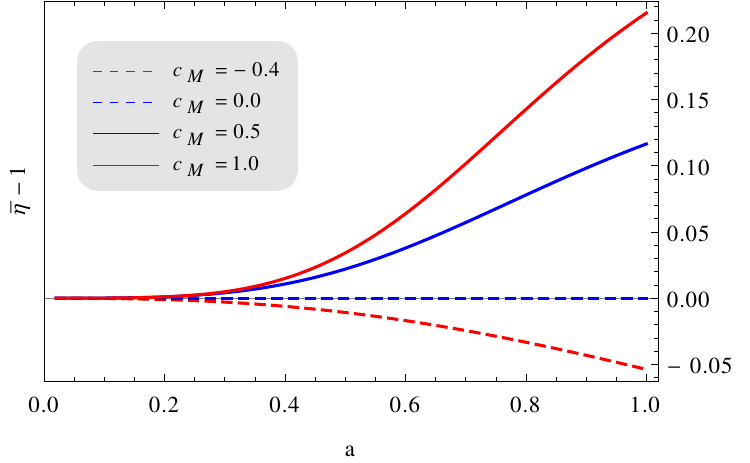}\\
\includegraphics[scale=0.42]{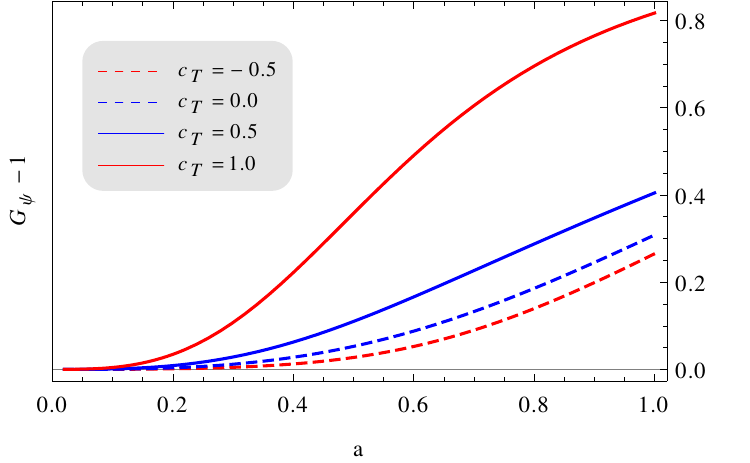}\includegraphics[scale=0.42]{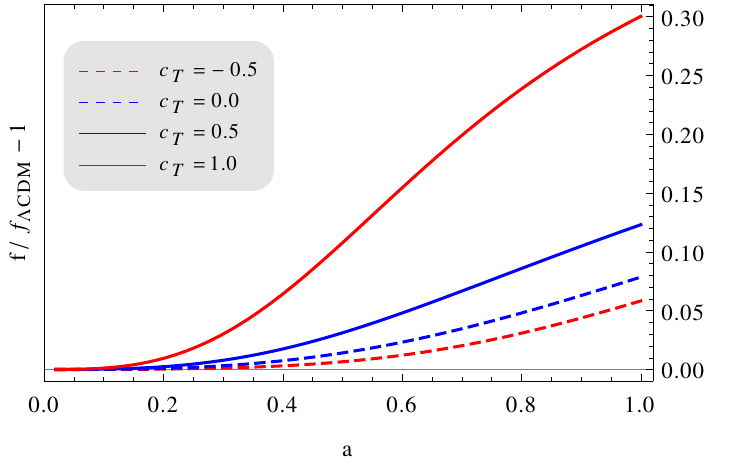}\includegraphics[scale=0.42]{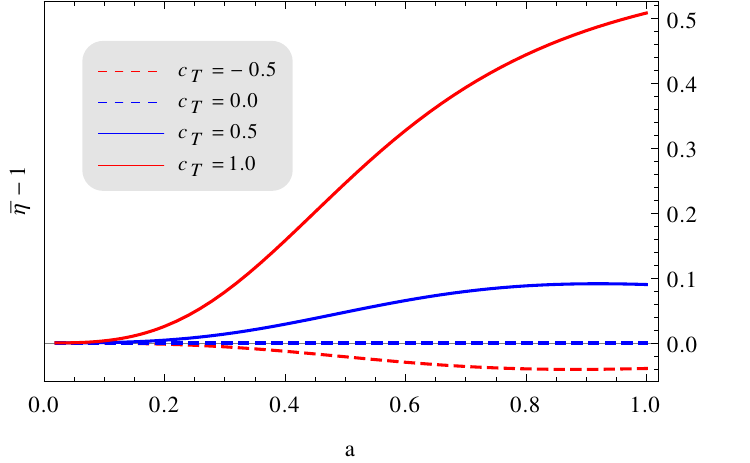}\protect\caption{Evolution of the effective Newton's constant $G_{\Psi}$ (left panels),
the linear growth rate $f$ (central panels) and the slip parameter
$\bar{\eta}$ (right panels) as a function of the scale factor $a$.
In every panel we take the relative deviations w.r.t.~our fiducial
model, i.e.~$\Lambda\textrm{CDM}$. In the top panels we fix $c_{\textrm{M}}=c_{\textrm{T}}=0$
and we vary $c_{\textrm{B}}=0.5$ (red dashed lines), $c_{\textrm{B}}=1.0$
(blue dashed lines), $c_{\textrm{B}}=1.5$ (blue solid lines) and
$c_{\textrm{B}}=2.0$ (red solid lines). In the central panels we
fix $c_{\textrm{T}}=0$, $c_{\textrm{B}}=1$ and we vary $c_{\textrm{M}}=-0.4$
(red dashed lines), $c_{\textrm{M}}=0$ (blue dashed lines), $c_{\textrm{M}}=0.5$
(blue solid lines) and $c_{\textrm{M}}=0.5$ (red solid lines). In
the bottom panels we fix $c_{\textrm{M}}=0$, $c_{\textrm{B}}=1$
and we vary $c_{\textrm{T}}=-0.5$ (red dashed lines), $c_{\textrm{T}}=0$
(blue dashed lines), $c_{\textrm{T}}=0.5$ (blue solid lines) and
$c_{\textrm{T}}=1$ (red solid line).\label{fig:noKfirst}}
\end{figure}

In Fig.~(\ref{fig:noKfirst}) we show the linear evolution for  few representative models. The left panels refer to the effective Newton's constant,
in the central panels we plot the linear growth rate and in the right
panels we show the time evolution of the slip parameter. In the top panels
 we present the behaviour of a minimally coupled theory ($c_{\textrm{M}}=c_{\textrm{T}}=0$).
We vary $c_{\textrm{B}}$ in the region where ghost and gradient instabilities
are avoided, i.e.~$c_{\textrm{B}}>0$. It is possible to note that
$c_{\textrm{B}}$ enhances the effective Newton's constant and the
linear growth. This is expected, and it is the signal of a stronger
gravity, which should be screened on small scales to satisfy solar
system constraints. It is also expected the absence of anisotropic
stress (top right panel), since $\bar{\eta}\neq1$ only in non-minimally
coupled theories. In the central panels we fix $c_{\textrm{B}}=1$
and $c_{\textrm{T}}=0$, letting $c_{\textrm{M}}$ free to vary. In
this case we can compensate the effect of the braiding by choosing
a negative value for $c_{\textrm{M}}$. By fixing $c_{\textrm{M}}\simeq-0.4$
we can recover nearly the standard predictions for $G_{\Psi}$, but
the growth rate $f$ appears slightly suppressed. This is because
in our description $c_{\textrm{M}}$ affects directly the evolution
of the matter density through Eq.~(\ref{eq:backdensity}). Finally,
in the bottom panels we show the evolution of $G_{\Psi}$, $f$ and
$\bar{\eta}$ for different values of $c_{\textrm{T}}$. We set $c_{\textrm{B}}=1$
and $c_{\textrm{M}}=0$. Note that we have chosen extreme values for
$c_{\textrm{T}}$, and $c_{\textrm{T}}=-0.5$ could be ruled out by
the Cherenkov radiation emission argument \cite{Moore:2001bv,Elliott:2005va,Kimura:2011qn}.
Despite this fact, the parameter $c_{\textrm{T}}$ seems to affect
the evolution of linear perturbations less than the others.

\begin{figure}
\centering\includegraphics[scale=0.6]{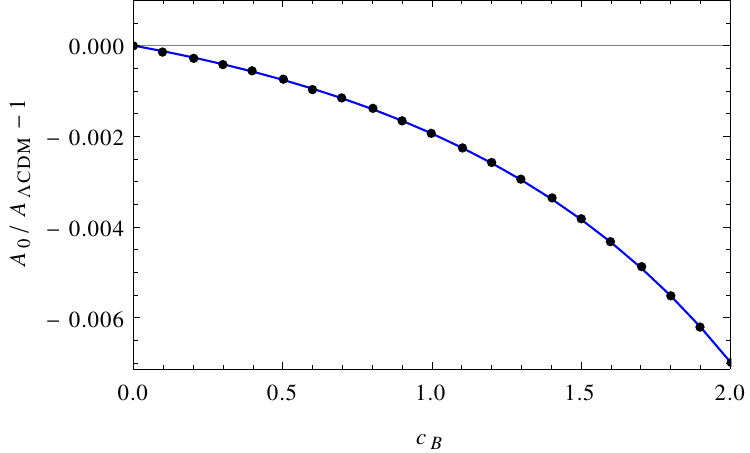}\includegraphics[scale=0.6]{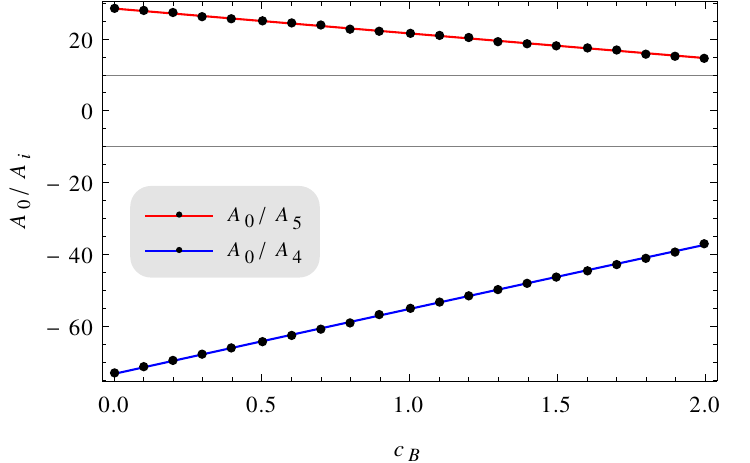}\\
\includegraphics[scale=0.6]{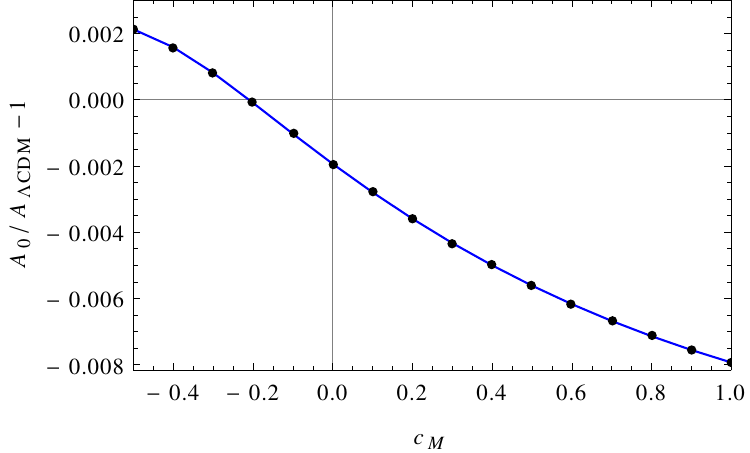}\includegraphics[scale=0.6]{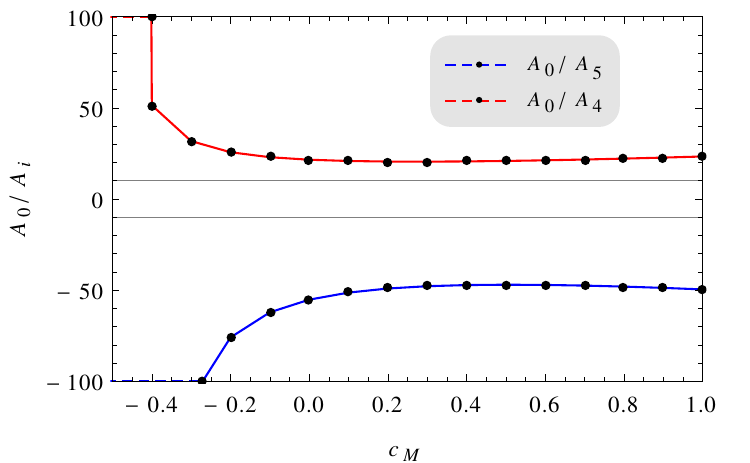}\\
\includegraphics[scale=0.6]{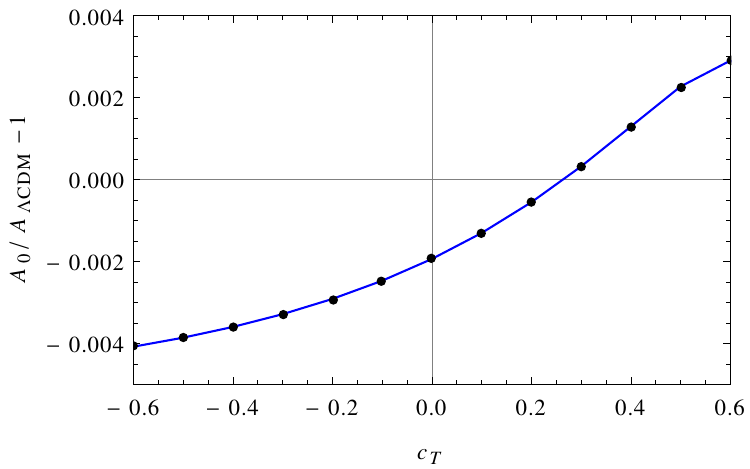}\includegraphics[scale=0.6]{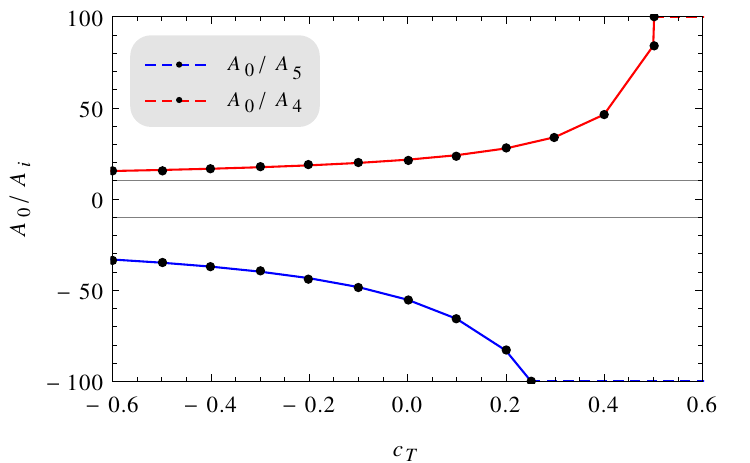}\protect\caption{In the left panels we show $A_{0}/A_{\Lambda\textrm{CDM}}$ evaluated
today, while in the right panels we show $A_{0}/A_{5}$ (red line)
and $A_{0}/A_{4}$ (blue line) evaluated today. In the top panels
we fix $c_{\textrm{M}}=c_{\textrm{T}}=0$ and we vary $c_{\textrm{B}}$.
In the central panels we fix $c_{\textrm{T}}=0$, $c_{\textrm{B}}=1$
and we vary $c_{\textrm{M}}$. In the bottom panels we fix $c_{\textrm{M}}=0$,
$c_{\textrm{B}}=1$ and we vary $c_{\textrm{T}}$. In the left panels
it is possible to note that the standard term of the kernel, Eq.~(\ref{eq:kernel}),
have deviations w.r.t.~the GR one $\lesssim1\%$ for every case under
consideration. The right panels show that $c_{5}$ and $c_{4}$ should
be $\gtrsim2$ in order to have an interesting effect on the bispectrum
kernel ($\sim10\%$). In the right central and bottom panels it is
possible to note divergencies in the curves for $c_{\textrm{M}}=-0.5$
and $c_{\textrm{T}}=0.6$ respectively. Indeed for these models $\left|A_{5},A_{4}\right|\ll A_{0}$,
which is a sign that these terms can not realistically modify the
bispectrum kernel.\label{fig:noKsecond}}
\end{figure}

As in the previous section, to  understand the bispectrum modifications we focus on the functions
$\{A_{0},A_{5},A_{4}\}$. In Fig.~(\ref{fig:noKsecond}) we show
the value assumed today by $A_{0}$ (left panels) and $A_{5}$ and
$A_{4}$ (right panels) as a function of $c_{\textrm{B}}$ (top panels),
$c_{\textrm{M}}$ (central panels) and $c_{\textrm{T}}$ (bottom panels).
For all the models under consideration $A_{0}$ deviates from its
standard value by less than $1\%$. Therefore also in this case   this
function cannot be responsible for large deviations in the bispectrum and only models with non-zero  $\alpha_5$ and $\alpha_{4}$ can modify it. 
Also the functions $A_{5}$ and $A_{4}$ appear rather suppressed
 compared to $A_{0}$. Indeed, $\left|A_{0}/A_{i}\right|>10$ for every
model under consideration. Moreover, for some value of $c_{\textrm{M}}$
(central right panel) and $c_{\textrm{T}}$ (bottom right panel) the
curves apparently diverge. For these models $\left|A_{5},A_{4}\right|\ll A_{0}$,
which indicates that these terms can not realistically modify the
bispectrum kernel. In the cases where $\left|A_{0}/A_{i}\right|$
is minimized we still need to choose $c_{5},c_{4}\gtrsim1$ in order
to achieve deviations in the kernel of at least $10\%$ . One could
argue that these deviation can be reached without imposing $c_{5},c_{4}\gtrsim1$
just by increasing the braiding $c_{\textrm{B}}$. However, comparing
these plots with the ones in Fig.~(\ref{fig:noKfirst}) we note that
the limit $c_{\textrm{B}}\gg1$ produces large modifications of the
first-order evolution. Even if we introduce a negative $c_{\textrm{M}}$
to compensate the linear-order deviations we expect for $A_{5}$ and
$A_{4}$ to be further suppressed (see central-right panel of Fig.~(\ref{fig:noKsecond})).

In conclusion, even for general models with large braiding,
at second-order in perturbation theory we can not produce large deviations
in the bispectrum kernel for models that respect the expected hierarchy,
Eq.~(\ref{eq:naturalness}) without generating even larger effects at first order.

\subsection{Bispectrum shapes}

In this section we show how the shape of the bispectrum is modified
for Horndeski-type DE/MG models. This is useful in order to estimate
what are the configurations for which the modifications of the bispectrum
kernel, if any, are maximised, i.e.,~the easiest to detect. In the
previous sections we considered the evolution of the function $C\left(t\right)$
and we showed that it is \textit{unnatural} to have large modifications
in $C\left(t\right)$ evaluated today. This is the only
modification that appears in the bispectrum kernel, Eq.~(\ref{eq:kernel}),
and, 
its magnitude at any particular time, is just a number.
 Therefore here in full generality we do not consider particular models, just the  value of  $C\left(z_{obs}\right)$, where $z_{obs}$ is the redshift
at which we want to measure the bispectrum. We  perturb  $C\left(z_{obs}\right)$ from it standard value 
($C_{\Lambda\textrm{CDM}}\simeq1.62$)
by fixed percentages ($1\%$, $2\%$, $5\%$, $10\%$). 
Recall that the deviations in the bispectrum kernel are maximised
today and vanish at early times by construction. However,  forthcoming and future surveys
do not measure the bispectrum today, but at higher redshifts ($z\sim0.5-1$).
Then, our main result, i.e.~the modifications of the bispectrum for
\textit{natural} models are $\lesssim1\%$ today, can be considered
as an upper bound on the modifications of the bispectrum that a generic
survey will measure if gravity satisfies our \textit{naturalness}
condition.

Instead of using the bispectrum itself, Eq.~(\ref{eq:bispectrumresult}),
we consider the reduced bispectrum \cite{Groth:1977gj, fryseldner}

\begin{equation}
Q\left(t,\vec{k}_{1},\vec{k}_{2},\vec{k}_{3}\right)\equiv\frac{B\left(t,\vec{k}_{1},\vec{k}_{2},\vec{k}_{3}\right)}{P\left(t,\vec{k}_{1}\right)P\left(t,\vec{k}_{2}\right)+cyc.}=\frac{F_{2}\left(t,\vec{k}_{1},\vec{k}_{2}\right)\,P\left(t,k_{1}\right)\,P\left(t,k_{2}\right)+cyc.}{P\left(t,\vec{k}_{1}\right)P\left(t,\vec{k}_{2}\right)+cyc.}\,,\label{eq:Q}
\end{equation}
which removes most of the cosmology and scale dependence. Here, the PS can be written as
\begin{equation}
P\left(t,\vec{k}\right)\propto\delta_{+}^{2}\left(t\right)\,T^{2}\left(k\right)\,\left(\frac{k}{H_{0}}\right)^{n_{s}},\label{eq:PS}
\end{equation}
where $\delta_{+}\left(t\right)$ is the growing solution of Eq.~(\ref{eq:diffdelta1})
and encodes all the time dependence of the PS. $T\left(k\right)$
is the transfer function, which we take as standard using the fitting
formula given in \cite{Bardeen:1985tr}. This is consistent with our
parametrization, since we only allow for modifications at late-times
while we assume that the Universe behaves as standard during the epochs
dominated by radiation and matter. Finally $n_{s}$ is the scalar
spectral index, and we fix it at $n_{s}=0.96$ \cite{Planck:2015xua}.
It is important to keep in mind that, since we are looking at models either
sub or super-braiding, the growth of perturbations is not scale dependent.
As a consequence, one can see using Eqs.~(\ref{eq:Q}-\ref{eq:PS}),
the only time dependence in the reduced bispectrum appears in the
kernel $F_{2}$, and thus in $C\left(t\right)$.

\begin{figure}
\centering\includegraphics[scale=0.65]{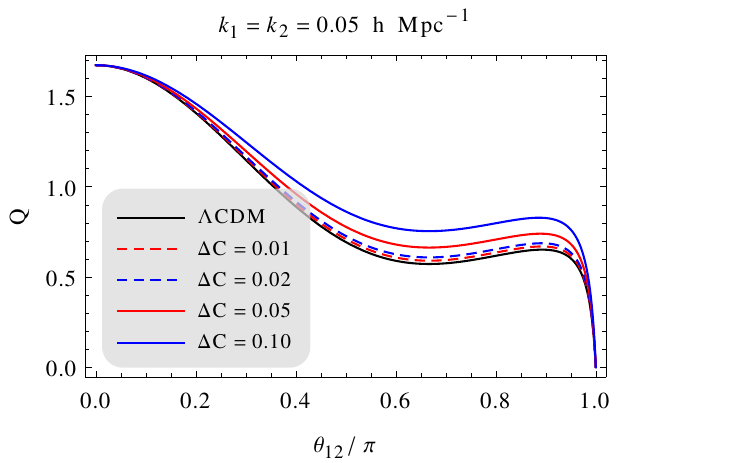}\includegraphics[scale=0.65]{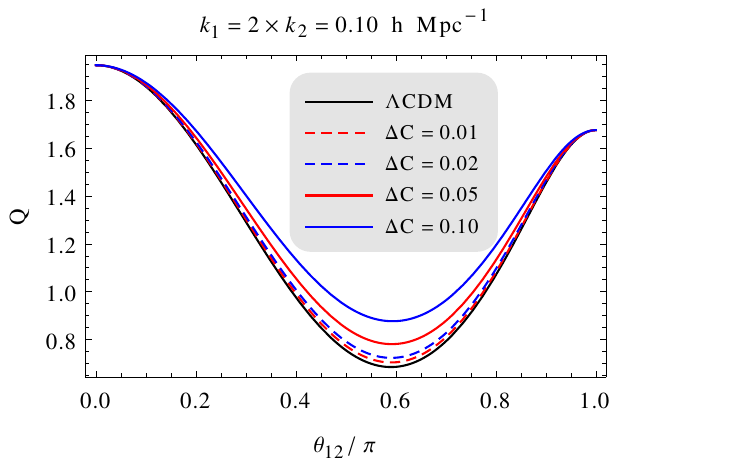}\\
\includegraphics[scale=0.65]{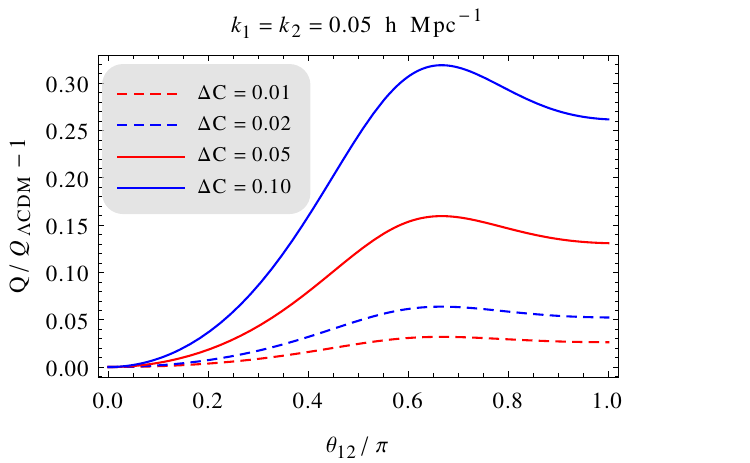}\includegraphics[scale=0.65]{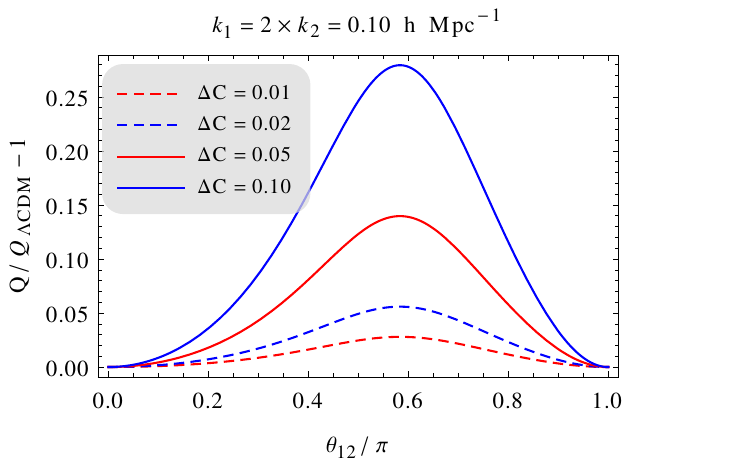}

\protect\caption{We show the reduced bispectrum $Q$ (top panels) and its relative
deviations w.r.t.~the reduced bispectrum of our fiducial model $Q_{\Lambda\textrm{CDM}}$
(bottom panels) for different configurations. In every panel $Q$
is plotted as a function of the angle $\theta_{12}\equiv\arccos\left(\mu\right)=\arccos\left(\frac{\vec{k}_{1}\cdot\vec{k}_{2}}{k_{1}k_{2}}\right)$.
In the left panels we fix $k_{1}=k_{2}=0.01\,h\,Mpc^{-1}$, in the
right panels $k_{1}=2\times k_{2}=0.02\,h\,Mpc^{-1}$. $\Delta C$
represents the relative modifications of the function $C$, i.e.~$\Delta C\equiv\frac{C-C_{\Lambda\textrm{CDM}}}{C_{\Lambda\textrm{CDM}}}$.
The models we plot are: $\Delta C=0.01$ (red dashed lines), $\Delta C=0.02$
(blue dashed lines), $\Delta C=0.05$ (red solid lines) and $\Delta C=0.10$
(blue solid lines). Note that in both panels the effects of a modified
bispectrum kernel are maximised for $\theta_{12}\simeq0.6\pi$, where
their magnitude reaches $\sim3\times\Delta C$.\label{fig:Q}}
\end{figure}
In Fig.~(\ref{fig:Q}), we show the shape
of the reduced bispectrum $Q$, Eq.~(\ref{eq:Q}), and its relative deviations w.r.t.~the reduced
bispectrum of our fiducial model ($Q_{\Lambda\textrm{CDM}}$) as a
function of the angle $\theta_{12}\equiv\arccos\left(\mu\right)=\arccos\left(\frac{\vec{k}_{1}\cdot\vec{k}_{2}}{k_{1}k_{2}}\right)$.
In the left panels we fix $k_{1}=k_{2}=0.05\,h\,Mpc^{-1}$, while in
the right panels we fix $k_{1}=2\times k_{2}=0.10\,h\,Mpc^{-1}$. The
different curves show the behaviour of $Q$ for different values of
$\Delta C$, which represents the relative modifications of the function
$C$ w.r.t.~our fiducial model, i.e.~$\Delta C\equiv\frac{C-C_{\Lambda\textrm{CDM}}}{C_{\Lambda\textrm{CDM}}}$.
It is important to note that in the left panel the limit $\theta_{12}\rightarrow\pi$ involve extremely large scales ($k_{i}\rightarrow0$).
Then, since we are assuming to be well inside or outside the braiding
scale, Eq.~(\ref{eq:braidingscale}), the behaviour of the curves
in these limits could probably be corrected for real models.

In every panel the differences in the reduced bispectrum are maximised at $\theta_{12}\simeq0.6\pi$, where they are three times bigger than the differences in $C$, i.e.~$\Delta Q\sim 3\Delta C$. For all the other configurations the deviations  from the standard bispectrum are suppressed. This can be connected with the findings of the previous sections, where we considered modifications in $C$ of $\sim3\%$ as large ones. Indeed, the observable is $Q$ in particular configurations, and not $C$. Then, we have been conservative in considering the maximum deviations as the deviations that could be detected by forthcoming and future surveys. Comparing Fig.~(\ref{fig:Q}) with the reduced bispectrum obtained by considering bias or non-linear corrections (e.g.~\cite{Bernardeau:2001qr}), we see that the Horndeski modifications of the DM bispectrum kernel, Eq.~(\ref{eq:kernel}),  are qualitatively different.

\section{Conclusions\label{sec:Conclusions}}

We have presented a systematic analysis of the DM bispectrum
generated at late-times by gravitational instability in the general
Horndeski class of models. We assumed a $\Lambda\textrm{CDM}$ evolution
for the Hubble function $H\left(t\right)$ and we parametrized the
perturbations. In this way we were able to separate the effects of
the expansion history of the Universe from the first and second-order
effects.

We focused on two class of models, the ones with large braiding ($\alpha_{\textrm{B}}^{2}\gg\alpha_{\textrm{K}}$)
and the ones with small braiding ($\alpha_{\textrm{B}}^{2}\ll\alpha_{\textrm{K}}$).
In the intermediate regime, where $\alpha_{\textrm{B}}^{2}\sim\alpha_{\textrm{K}}$,
we expect the linear evolution of perturbations to be scale-dependent.
Then, any deviation from standard gravity can be seen looking, e.g., ~at the
PS without the need of a bispectrum analysis.

In the small braiding case we noted that the only parameter that is
left free to vary at linear-order is the Planck mass run-rate ($c_{\textrm{M}}$).
Then, we showed that, even if we allow for large deviations at linear-order
$\sim10\%$ ($\sim1\%$) we can not have large modifications at second-order
$\sim3\%$ ($\sim1\%$) if $c_{5},c_{4}\lesssim c_{\textrm{M}},c_{\textrm{T}}$. Since the parameters $c_{5}$
and $c_{4}$ represent the second-order contribution of non minimal
coupling between the metric $g_{\mu\nu}$ and the scalar field $\phi$,
they can be seen as corrections of the linear parameters $c_{\textrm{M}}$
and $c_{\textrm{T}}$. This argument leads to Eq.~(\ref{eq:naturalness}),
which we consider to be the natural behaviour of realistic DE/MG models.

The large braiding case is more complicated. We then explored the parameter
space for realistic values of $\{c_{\textrm{B}},c_{\textrm{M}},c_{\textrm{T}}\}$.
We found that, assuming that the linear growth $f$ is measured with a  $6\%$ precision (or better), there
can not be significant deviations in the bispectrum if $c_{5},c_{4}\lesssim c_{\textrm{M}},c_{\textrm{T}}$.
The only way of having large modifications in the bispectrum kernel
($\sim3\%$) with $c_{5},c_{4}\lesssim c_{\textrm{M}},c_{\textrm{T}}$, is to allow for unrealistic
deviations on the growth rate $f$ ($\sim30\%$). These models are expected to be excluded by linear
observations, without the need of the analysis of the bispectrum.

Large and observable modifications of the DM bispectrum kernel are still possible in the Horndeski class of models. However, they can not come from the function $A_0$ in Eq.~(\ref{eq:A0A5A6}), since it can be modified by $\lesssim 1\%$ for all the models under consideration. Indeed, only models with with large $\alpha_5$ and $\alpha_{4}$, i.e.~$\alpha_{5},\alpha_{4}\gg\alpha_{\textrm{M}},\alpha_{\textrm{T}}$, have a chance to modify it. Note that  in popular MG models such as  as Brans-Dicke/f(R), kinetic gravity braiding, cubic galileons these functions are zero. It is therefore not surprising that no significant  signatures of these models in the bispectrum  were found in these cases. Large  $\alpha_5$,  $\alpha_{4}$ imply a large non-minimal coupling between the curvature and the derivatives of the scalar field.\footnote{As suggested by non-perturbative analysis of kinetic mixing in scalar-tensor theories \cite{Bettoni:2015wta}.} This can be achieved in, e.g., quartic and quintic galileons at the price of having large $\alpha_{\textrm{M}}$ and $\alpha_{\textrm{T}}$.

In conclusion, we have shown that it is not possible
to have large modifications in the DM bispectrum kernel for models following our parameterisation of the free Horndeski functions and 
that satisfy our \textit{naturalness} condition, Eq.~(\ref{eq:naturalness}), without introducing even larger changes in first-order quantities.
Even if in these models there is no extra qualitatively new information
in the DM bispectrum, we are  certainly not advocating not to use it. Our findings imply that 
the kernel, Eq.~(\ref{eq:kernel}), can be modelled as the standard
GR one  and applied more generally. Such  bispectrum analysis is still useful to: 1) have new information on real word effects, e.g.,~bias parameters, to remove degeneracies;
2) decrease statistical errors; 3) offer a powerful consistency check.

Eq.~(\ref{eq:naturalness}) can be seen as a prescription, under which we can consider the DM bispectrum kernel as standard. Then, the observation of a large bispectrum deviation
that can not be explained in terms of bias would imply either that the linear evolution of perturbations is strongly different than the
evolution predicted by GR or that the theory of gravity is exotic and/or fine-tuned.

\section*{Acknowledgements}

The work of EB and LV is supported by the European Research Council
under the European Communityâ Seventh Framework Programme FP7- IDEAS-Phys.LSS
240117.  LV is supported in part by Mineco grant FPA2011-29678- C02-02. The authors are grateful to Antonio J.~Cuesta, Cristiano Germani, Ignacy Sawicki,
Fergus Simpson and Miguel Zumalacarregui for valuable comments and
criticisms.

\appendix

\section{Formulas\label{sec:Formulas}}

In this Section we present the equations that are too long to fit
in the main text. In particular, the second-order $\alpha_{i}$ functions
in Sec.~\ref{sec:Approach} read

\begin{align}
M_{*}^{2}\alpha_{5}\equiv & 2\dot{\phi}HXG_{5X}\label{eq:alpha5}\\
M_{*}^{2}\alpha_{4}\equiv & -2X\left(G_{4X}+2XG_{4XX}-G_{5\phi}-XG_{5\phi X}\right)-2\dot{\phi}HX^{2}G_{5XX}\,,\label{eq:alpha6}
\end{align}
where the functions $G_{4,5}$ are the Horndeski functions and their
derivatives defined in Eq.~(\ref{eq:action}).

The modified Newton's constants in Eqs.~(\ref{eq:Poissonpsi}-\ref{eq:Poissonphi})
are 
\begin{align}
G_{\Psi}= & 1-\frac{\left(\alpha_{\textrm{B}}+2\alpha_{\textrm{M}}\right)^{2}H^{2}-\alpha_{\textrm{T}}\left[2\left(2\dot{H}+\tilde{\rho}_{\textrm{m}}\right)-2\left(H\alpha_{\textrm{B}}\right)^{.}+H^{2}\left(2-\alpha_{\textrm{B}}\right)\left(\alpha_{\textrm{B}}+2\alpha_{\textrm{M}}\right)\right]}{2\left(2\dot{H}+\tilde{\rho}_{\textrm{m}}\right)-2\left(H\alpha_{\textrm{B}}\right)^{.}-H^{2}\left(2-\alpha_{\textrm{B}}\right)\left[\alpha_{\textrm{B}}+2\alpha_{\textrm{M}}-\alpha_{\textrm{T}}\left(2-\alpha_{\textrm{B}}\right)\right]}\label{eq:Gpsi}\\
G_{\Phi}= & 1-\frac{\alpha_{\textrm{B}}H^{2}\left[\alpha_{\textrm{B}}+2\alpha_{\textrm{M}}-\alpha_{\textrm{T}}\left(2-\alpha_{\textrm{B}}\right)\right]}{2\left(2\dot{H}+\tilde{\rho}_{\textrm{m}}\right)-2\left(H\alpha_{\textrm{B}}\right)^{.}-H^{2}\left(2-\alpha_{\textrm{B}}\right)\left[\alpha_{\textrm{B}}+2\alpha_{\textrm{M}}-\alpha_{\textrm{T}}\left(2-\alpha_{\textrm{B}}\right)\right]}\,.\label{eq:Gphi}
\end{align}

At second-order, the function $C\left(t\right)$ that modifies the
bispectrum kernel, Eq.~(\ref{eq:kernel}), reads
\begin{equation}
C\left(t\right)\equiv\int_{0}^{t}\frac{dt^{\prime}\delta_{g}^{2}\left(t^{\prime}\right)}{W\left(t^{\prime}\right)\delta_{g}^{2}\left(t\right)}\left[\delta_{d}\left(t\right)\delta_{g}\left(t^{\prime}\right)-\delta_{g}\left(t\right)\delta_{d}\left(t^{\prime}\right)\right]c\left(t^{\prime}\right)\,,\label{eq:C}
\end{equation}
where $\delta_{g}$ and $\delta_{d}$ are the growing and the decaying modes of Eq.~(\ref{eq:diffdelta1}) respectively, while $W\left(t\right)\equiv\delta_{g}\left(t\right) \delta^\prime_{d}\left(t\right)- \delta^\prime_{g} \delta_{d}\left(t\right)\left(t\right)$ is the Wronskian. Finally $c\left(t\right)$ is defined as
\begin{align}
c\left(t\right)= & \tilde{\rho}_{\textrm{m}}G_{\Psi}+\frac{8}{3}H^{2}f^{2}+\frac{\tilde{\rho}_{\textrm{m}}^{2}}{3H^{2}\alpha_{\textrm{B}}^{2}}\left(3+2\alpha_{\textrm{T}}-3G_{\Psi}\right)\left(1-G_{\Phi}\right)^{2}\label{eq:c}\\
 & +\frac{2\tilde{\rho}_{\textrm{m}}^{2}\left(1+\alpha_{\textrm{T}}-G_{\Psi}\right)\left(1-G_{\Phi}\right)}{3H^{2}\alpha_{\textrm{B}}^{2}\left[\alpha_{\textrm{B}}\left(1+\alpha_{\textrm{T}}\right)+2\left(\alpha_{\textrm{M}}-\alpha_{\textrm{T}}\right)\right]}\left[\left(\alpha_{\textrm{B}}-2\alpha_{\textrm{M}}\right)\left(1-G_{\Phi}\right)-2\alpha_{\textrm{B}}\alpha_{\textrm{T}}\right]\nonumber \\
 & -\frac{4\tilde{\rho}_{\textrm{m}}^{2}\left(1+\alpha_{\textrm{T}}-G_{\Psi}\right)\left(1-G_{\Phi}\right)^{2}\dot{\alpha}_{4}}{3H^{3}\alpha_{\textrm{B}}^{2}\left[\alpha_{\textrm{B}}\left(1+\alpha_{\textrm{T}}\right)+2\left(\alpha_{\textrm{M}}-\alpha_{\textrm{T}}\right)\right]}-\frac{2\tilde{\rho}_{\textrm{m}}^{2}\left(1-G_{\Phi}\right)^{2}G_{\Psi}\alpha_{4}}{3H^{2}\alpha_{\textrm{B}}^{2}}\nonumber \\
 & +\frac{4\tilde{\rho}_{\textrm{m}}^{2}\left(1+\alpha_{\textrm{T}}-G_{\Psi}\right)\left(1-G_{\Phi}\right)\left[\left(1-G_{\Phi}\right)\left(1-\alpha_{\textrm{M}}\right)-\alpha_{\textrm{B}}G_{\Psi}\right]\alpha_{4}}{3H^{2}\alpha_{\textrm{B}}^{2}\left[\alpha_{\textrm{B}}\left(1+\alpha_{\textrm{T}}\right)+2\left(\alpha_{\textrm{M}}-\alpha_{\textrm{T}}\right)\right]}\nonumber \\
 & +\frac{\tilde{\rho}_{\textrm{m}}^{2}\left(1+\alpha_{\textrm{T}}-G_{\Psi}\right)\left(1-G_{\Phi}\right)\left[2\left(1-G_{\Phi}\right)+\alpha_{\textrm{B}}\left(1-3G_{\Phi}\right)\right]}{3H^{2}\alpha_{\textrm{B}}^{2}\left[\alpha_{\textrm{B}}\left(1+\alpha_{\textrm{T}}\right)+2\left(\alpha_{\textrm{M}}-\alpha_{\textrm{T}}\right)\right]}\left(\frac{\alpha_{5}}{H}\right)^{.}\nonumber \\
 & -\frac{\tilde{\rho}_{\textrm{m}}^{2}\left(1-G_{\Phi}\right)^{2}}{3\alpha_{\textrm{B}}^{2}H^{2}}\left(\frac{\alpha_{5}}{H}\right)^{.}+\frac{\tilde{\rho}_{\textrm{m}}^{2}\left(1-G_{\Phi}\right)^{2}\left(1+2G_{\Psi}-\alpha_{\textrm{M}}\right)}{3\alpha_{\textrm{B}}^{2}H^{2}}\alpha_{5}\nonumber \\
 & -\frac{\alpha_{5}\tilde{\rho}_{\textrm{m}}^{2}G_{\Psi}}{3\alpha_{\textrm{B}}H^{2}}\left(1-G_{\Phi}^{2}\right)+\frac{8\dot{H}\tilde{\rho}_{\textrm{m}}^{2}\left(1+\alpha_{\textrm{T}}-G_{\Psi}\right)\left(1-G_{\Phi}\right)^{2}\alpha_{5}}{3H^{4}\alpha_{\textrm{B}}^{2}\left[\alpha_{\textrm{B}}\left(1+\alpha_{\textrm{T}}\right)+2\left(\alpha_{\textrm{M}}-\alpha_{\textrm{T}}\right)\right]}\nonumber \\
 & -\frac{\tilde{\rho}_{\textrm{m}}^{2}\left(1+\alpha_{\textrm{T}}-G_{\Psi}\right)\left[2\left(1-G_{\Phi}\right)^{2}\left(1-\alpha_{\textrm{M}}\right)-\left(1-2G_{\Phi}\right)\alpha_{\textrm{B}}^{2}G_{\Psi}\right]\alpha_{5}}{3H^{2}\alpha_{\textrm{B}}^{2}\left[\alpha_{\textrm{B}}\left(1+\alpha_{\textrm{T}}\right)+2\left(\alpha_{\textrm{M}}-\alpha_{\textrm{T}}\right)\right]}\nonumber \\
 & -\frac{\tilde{\rho}_{\textrm{m}}^{2}\left(1+\alpha_{\textrm{T}}-G_{\Psi}\right)\left(1-G_{\Phi}\right)\left[\left(1-3G_{\Phi}\right)\left(1-\alpha_{\textrm{M}}\right)-4G_{\Psi}\right]\alpha_{5}}{3H^{2}\alpha_{\textrm{B}}\left[\alpha_{\textrm{B}}\left(1+\alpha_{\textrm{T}}\right)+2\left(\alpha_{\textrm{M}}-\alpha_{\textrm{T}}\right)\right]}\,.\nonumber 
\end{align}

\bibliographystyle{utcaps}
\bibliography{biblio}

\end{document}